\documentclass[conference]{IEEEtran}
\usepackage{bbm,dsfont,mathrsfs,fixmath}
\usepackage{stmaryrd}
\usepackage{amsmath,amstext,amsfonts,amssymb}
\usepackage{epsfig,float}
\usepackage{upgreek}
\usepackage{graphics}
\usepackage{ dsfont }
\usepackage{cite}
\usepackage{psfrag}
\usepackage{subfigure}
\usepackage{url}
\usepackage{shadow,color,pifont,times,rotate}

\newcommand{\mc}{\mathcal}

\def\esp{{\mathbb E}}  
 \DeclareMathAlphabet{\mathpzc}{OT1}{pzc}{m}{it}
\newtheorem{theorem}{Theorem}

\newtheorem{definition}{Definition}

\usepackage{enumerate}    
\ifCLASSINFOpdf
		\usepackage{epstopdf}
\else
\fi
\usepackage{enumerate}    
\DeclareRobustCommand{\prob}[1][P]{\ensuremath {\mathbb{#1}}}

\IEEEoverridecommandlockouts
\begin{document}
\title{Selective Coding Strategy for Unicast \\Composite Networks}

\author{\IEEEauthorblockN{Arash Behboodi}
\IEEEauthorblockA{Dept. of Telecommunications, SUPELEC\\
91192 Gif-sur-Yvette, France\\
Email: \{arash.behboodi\}@supelec.fr}

\and
\IEEEauthorblockN{Pablo Piantanida}
\IEEEauthorblockA{Dept. of Telecommunications, SUPELEC\\
91192 Gif-sur-Yvette, France\\
Email: \{pablo.piantanida\}@supelec.fr}

\thanks{The work of P. Piantanida is partially supported by the ANR grant (FIREFLIES) INTB 0302 01.\vspace{-2mm}}}
\maketitle

\begin{abstract}
Consider a composite unicast relay network where the channel statistic is randomly drawn from a set of conditional distributions indexed by $\theta\in\Theta$, which is assumed to be unknown at the source, fully known at the destination and only partly known at the relays. Commonly, the coding strategy at each relay is fixed regardless of its channel measurement. A novel coding for unicast composite networks with multiple relays is introduced. This enables the relays to select dynamically --based on its channel measurement-- the best coding scheme between compress-and-forward (CF) and decode-and-forward (DF). As a part of the main result, a  generalization of Noisy Network Coding  is shown for the case of unicast general networks where the relays are divided between those  using DF  and CF coding. Furthermore, the relays using DF scheme can exploit the help of those based on CF scheme via offset coding. It is demonstrated via numerical results that this novel coding, referred to as Selective Coding Strategy (SCS), outperforms conventional coding schemes.  
\end{abstract}
%\IEEEpeerreviewmaketitle

\section{Introduction}
Multiterminal networks are the essential part of modern telecommunication systems. Recently, these were studied from various aspects. The cutset bound for the general multicast network was established in \cite{Elias1956}. Network coding theorem for the graphical multicast network was investigated in \cite{Ahlswede2000} where the max-flow min-cut theorem for network information flow was presented for the point-to-point communication network. Whereas Lim \textit{et al.} proposed the Noisy Network Coding (NNC) scheme for the general multicast network, which includes most of the existing bounds on multiterminal networks\cite{Lim2011}. Kramer \textit{et al.} developed an inner bound for a point-to-point general network using decode-and-forward (DF) which achieves the capacity of the degraded multicast network \cite{Kramer2005}. 

For the above mentioned scenarios, the probability distribution (PD) of the network is supposed to be fixed during the communication and hence available to all nodes beforehand. However, wireless channels are essentially time-varying due to fading and user mobility, and hence the terminals do not have full knowledge of all channel parameters involved in the communication. In particular,  without feedback channel state information (CSI) cannot be available to the encoders end. During years, an ensemble of research activities has been dedicated to both theoretical and practical aspects of communication in presence of channel uncertainty. From an information-theoretic viewpoint, the compound channel, first introduced by Wolfowitz \cite{Wolfowitz1960} is one of the most important models that deals with channel uncertainty, and continues to attract much attention from researchers (see \cite{Lapidoth1998A}  and references therein). Composite models are more appropriate to deal with wireless scenarios since unlike compound models they deal with channel uncertainty by introducing a PD $\prob_{\uptheta}$ on the channel selection. These models consist of a set of conditional PDs from which the current channel index $\theta$, which can be a vector of parameters, is drawn according to $\prob_{\uptheta}$ and remains fixed during the communication. An example of this model can be slowly fading channels. Capacity for this class of channels has been widely studied beforehand (see \cite{Effros2010} and references therein), for wireless scenarios via the well-known notion of outage capacity (see \cite{720551} and references therein) and oblivious cooperation over fading Gaussian channels 
in \cite{Katz2009, Behboodi2010A, Behboodi2011A}.  

In this paper, we study the composite multiple relay network where the channel index $\theta\in\Theta$ is randomly drawn according to $\prob_{\uptheta}$. The index $\theta=(\theta_r,\theta_d)$ remains fixed during the communication but is unknown at the source, fully known at the destination and partly known $\theta_r$ at the relays end. Although a compound approach can guarantee asymptotically zero-error probability regardless of $\theta$, it would be not an adequate choice for most of wireless models.  As a different approach, the coding rate $r$ is selected regardless of the current index. Hence the encoder cannot necessarily guarantee arbitrary small error probability. In this case the asymptotic error probability becomes the measure of interest, characterizing the reliability function. Moreover, it turns out that depending on the channel draw, there may not be a unique set of relay functions that minimizes the error probability. In other words, the relay function should be chosen based on the channel parameters. However, since full CSI is not available to all nodes, the relay functions are usually chosen regardless of their channel measurements which becomes the bottleneck of the coding. We present a novel coding strategy from which the relays can select, based on their measurements, an adequate coding strategy. To this purpose an achievable region that generalizes NNC to the case of mixed coding strategy, where DF relays exploit the help of CF relays, is derived. 

Section II presents definitions and Section III introduces main results, and the sketch of proofs is relegated to Section IV. Finally, numerical evaluation over the slow-fading Gaussian two relay channel is presented in Section V.  
%%%%%%%%%%%%%%%%%%%%%%%%%%%%%%%%%%%%%%%%%%%%%%%%%%%%%%%%%%%%%%%%%%%%%%%%%%%%%%%%%%%%%%%%%%%%%%%%%%%%%%%%%%%%%%%%%%%%%%%%%%%%%%%%%%%%
%%%%%%%%%%%%%%%%%%%%%%%%%%%%%%%%%%%%%%%%%%%%%%%%%%%%%%%%%%%%%%%%%%%%%%%%%%%%%%%%%%%%%%%%%%%%%%%%%%%%%%%%%%%%%%%%%%%%%%%%%%%%%%%%%%%% 
%% DEFINITIONS
%%%%%%%%%%%%%%%%%%%%%%%%%%%%%%%%%%%%%%%%%%%%%%%%%%%%%%%%%%%%%%%%%%%%%%%%%%%%%%%%%%%%%%%%%%%%%%%%%%%%%%%%%%%%%%%%%%%%%%%%%%%%%%%%%%%% %%%%%%%%%%%%%%%%%%%%%%%%%%%%%%%%%%%%%%%%%%%%%%%%%%%%%%%%%%%%%%%%%%%%%%%%%%%%%%%%%%%%%%%%%%%%%%%%%%%%%%%%%%%%%%%%%%%%%%%%%%%%%%%%%%%% 
\section{Problem Definition }
\begin{figure} [t]
\centering  
\includegraphics [width=.45 \textwidth] {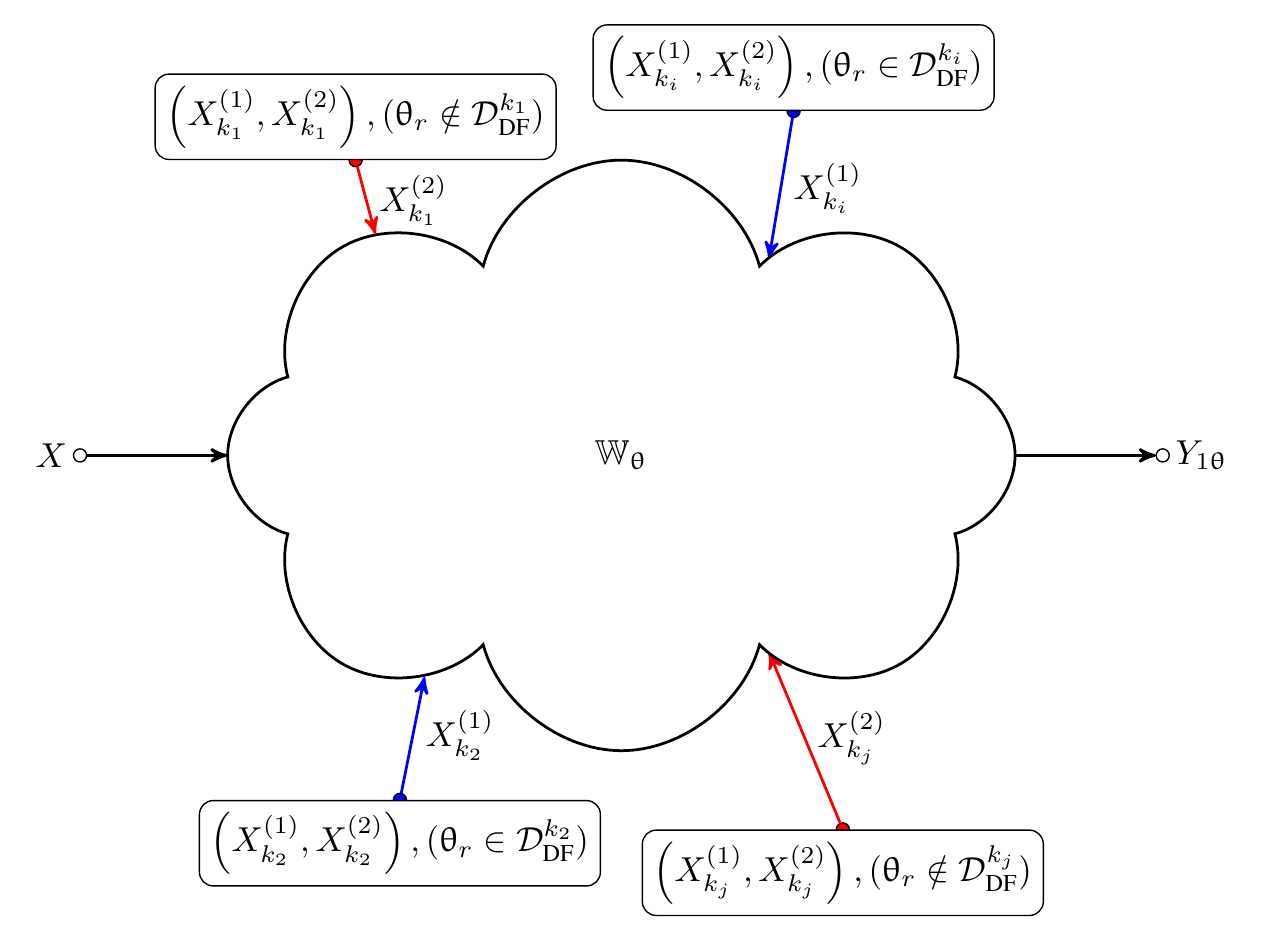}    
\caption{Composite Multiple Relay Network.}
\label{fig:CMRN}
\end{figure}

The composite multiple relay channel consists of a set of multiple relay channels, as depicted in Fig. \ref{fig:CMRN} and denoted by
$$
\big\{\prob[W]^n_\theta=P_{Y_{1,\theta}^{n} Z_{1\theta_r}^{n}Z_{2\theta_r}^{n}\dots Z_{N\theta_r}^{n}|X^{n} X^{n}_{1\theta_r} X^{n}_{2\theta_r} \dots X^{n}_{N\theta_r}}\big \}^{\infty}_{n=1}
$$ 
where $X$ denotes the channel input, $X_{k\theta_r}$ and $Z_{k\theta_r}$ the relay inputs and outputs and $Y_{1\theta}$ the channel output. We assume a memoryless multiple relay channel with $N$ relays but single  source and destination. The channels are indexed by vector of parameters $\theta=(\theta_d,\theta_r)$ with $\theta_d\in\Theta_d,\theta_r \in {\Theta_r}$, where $\theta_r$ denotes all parameters affecting the relays' output and $\theta_d$ are the remaining parameters involved in the communication. Let $\prob_\uptheta$ be a joint probability measure on ${\Theta}$. 
The channel parameters affecting relay and destination outputs $\uptheta=(\uptheta_r,\uptheta_d)$ are drawn according to the joint PD $\prob_{\uptheta}$ and remain fixed during the communication. However, the specific draw of $\uptheta$ is assumed to be unknown at the source, fully known at the destination and partly known $\uptheta_r$ at the relays end. Assume that $\mc{N}=\{1,\dots,N\}$ and for any $\mc{S}\subseteq\mc{N}$, $X_{\mc{S}}=\{X_i:\,\,i\in \mc{S}\}$. \vspace{2mm}

\begin{definition}[code and achievability] \label{def-code}
A code-$\mc{C}(n,M_n,r)$ for the composite multiple relay channel consists of:
\begin{itemize}
\item An encoder mapping  $\left \{\varphi:\mc{M}_{n} \longmapsto \mathcal{X}^n  \right\}$, 
\item A decoder mapping $\left\{ \phi_{\theta}:\mathcal{Y}_1^n \longmapsto \mc{M}_{n}\right\}$,
\item A set of relay functions $\left\{ f^{(k)}_{i{\theta_r}} :\mathcal{Z}_k^{i-1}  \longmapsto \mathcal{X}_{k} \right \}_{i=1}^n$ for $k\in\mc{N}$. Only partial CSI at the relay  is assumed which is mainly related to the $k$-th source-to-relay channel. 
\end{itemize}
An error probability $0\leq \epsilon<1$  is said to be $r$-achievable, if there exists a code-$\mc{C}(n,M_n,r)$ with rate satisfying
\begin{equation}
\liminf\limits_{n\rightarrow \infty}\frac{1}{n} \log M_n \geq r 
\end{equation} 
and average error probability 
\begin{equation}
 \limsup\limits_{n\rightarrow \infty}\, \esp_{\uptheta}\left[\Pr\left\{\phi_{\uptheta}(Y^n_{1\uptheta}) \neq W\big | \uptheta\right\}\right]\leq \epsilon.
\label{errorprob_def}   
\end{equation}    
The infimum of all $r$-achievable EPs $\bar{\epsilon}(r)$ is defined as 
\begin{equation}
\bar{\epsilon}(r)=\inf\left\{0\leq \epsilon<1\,:\, \textrm{$\epsilon$ is $r$-achievable}\right\}.
\label{errorprob_def2}   
\end{equation}    
We emphasize that for channels satisfying the strong converse property and with unique best code word, e.g., Gaussian slow-fading single user channel, \eqref{errorprob_def2} coincides with the definition of the outage probability for the unique best codeword.
\end{definition}
In the present setting, we assume that the source is not aware of the specific draw $\theta\sim\prob_{\uptheta }$ and hence, the coding rate $r$ and the coding strategy --DF or CF scheme-- must be chosen independent of the channel draw. Furthermore, both remain fixed during the communication  regardless of the channel measurement at the relays end. We aim to characterize the smallest possible average error probability as defined by \eqref{errorprob_def}, as a function of the coding rate $r$. 

It can be shown that $\bar{\epsilon}(r)$ can be bounded as follows
\begin{align}
\prob_\uptheta(r \in \mc{S}_\theta) \leq \bar{\epsilon}(r) \leq \inf_{\mc{C}}\prob_\uptheta(r \notin\mc{R}_\theta(\mc{C})),
\label{general-bounds}   
\end{align}
% 
% \begin{equation}
% \prob_{\uptheta }({r}>{S}_{\uptheta}) \leq \bar{\epsilon}(r) \leq \prob_{\uptheta }( r >{C}_{\uptheta}),
% \label{general-bounds}   
% \end{equation}  
where $\mc{R}_\theta$ is any achievable rate for the unicast network with a given $\theta$, and $\mc{S}_\theta$ is 
% the full error region of this channel for a given $\theta$. 
% where ${C}_{\uptheta }$ is the capacity (unknown) of the unicast network and ${S}_\uptheta$ for fixed $\uptheta$ is
the infimum of all rates  such that every code with such rate yields error probability tending to one, and $\mc{C}$ as all codes. It can be shown that ${S}_\uptheta$ can be replaced with max-flow min-cut bound. A special case of this result has been proved in recent work \cite{Behboodi2011A} for the relay channel. 
% Obviously, the capacity can be replaced by any achievable rate which still provides an upper bound on $\bar{\epsilon}(r)$.
%%%%%%%%%%%%%%%%%%%%%%%%%%%%%%%%%%%%%%%%%%%%%%%%%%%%%%%%%%%%%%%%%%%%%%%%%%%%%%%%%%%%%%%%%%%%%%%%%%%%%%%%%%%%%%%%%%%%%%%%%%%
%%%%%%%%%%%%%%%%%%%%%%%%%%%%%%%%%%%%%%%%%%%%%%%%%%%%%%%%%%%%%%%%%%%%%%%%%%%%%%%%%%%%%%%%%%%%%%%%%%%%%%%%%%%%%%%%%%%%%%%%%%%
%% Composite Multiple Relay Networks
%%%%%%%%%%%%%%%%%%%%%%%%%%%%%%%%%%%%%%%%%%%%%%%%%%%%%%%%%%%%%%%%%%%%%%%%%%%%%%%%%%%%%%%%%%%%%%%%%%%%%%%%%%%%%%%%%%%%%%%%%%%
%%%%%%%%%%%%%%%%%%%%%%%%%%%%%%%%%%%%%%%%%%%%%%%%%%%%%%%%%%%%%%%%%%%%%%%%%%%%%%%%%%%%%%%%%%%%%%%%%%%%%%%%%%%%%%%%%%%%%%%%%%%
\section{Composite Multiple Relay Networks}

Consider the composite unicast network with multiple relays and parameters $\uptheta$. The rate is fixed to $r$ and so is the source code. The goal is to minimize the expected error probability. The common option is that each relay fixes its coding strategy, namely DF or Compress-and-Forward (CF), regardless of $\uptheta$. In other words, the relays with index in $\mc{V}\subseteq\mc{N}$ will use CF scheme. For instance, to evaluate the expected error probability we first present an achievable rate for the multiple relay network where NNC is generalized to networks with mixed cooperative strategy. Part of relays are using DF coding while the reminding relays use CF scheme. Moreover, DF relays exploit the help of CF relays to decode the source message. Using this theorem, an achievable rate can be obtained for every set $\mc{V}$ of relay nodes. \vspace{1mm}
\begin{theorem} [Cooperative Mixed NNC]
For the multiple relay channel, the following rate is achievable 
\begin{IEEEeqnarray}{ll}
R \leq \max\limits_{P\in\mc{P}} \max\limits_{\mc{V}\subseteq\mc{N}}\min  \Big\{&\max_{\mc{T}\in\Upsilon(\mc{V})}\min_{\mc{S}\subseteq\mc{T}}  R_{\mc{T}}(\mc{S}),\nonumber\\
&\min_{k\in\mc{V}^c}\max_{\mc{T}_k\in\Upsilon_k(\mc{V})}\min_{\mc{S}\subseteq\mc{T}_k} R^{(k)}_{\mc{T}_k}(\mc{S})\Big\}
\label{CMNNC}%\vspace{-2mm}
\end{IEEEeqnarray} 
with 
\begin{IEEEeqnarray*}{lCl}
R_{\mc{T}}(\mc{S})&=&I(XX_{\mc{V}^c}X_{\mc{S}};\hat{Z}_{\mc{S}^c}Y_1|X_{\mc{S}^c}Q)\\
&-&I(Z_{\mc{S}};\hat{Z}_{\mc{S}}|XX_{\mc{T}\cup \mc{V}^c}\hat{Z}_{\mc{S}^c}Y_1Q)\,\,\,\,\,\,\,\,\,\, (\mc{S}^c=\mc{T}-\mc{S}), \\
R^{(k)}_{\mc{T}_k}(\mc{S})&=&I(X;\hat{Z}_{\mc{T}_k}Z_k|X_{\mc{V}^c}X_{\mc{T}_k}Q)+I(X_{\mc{S}};Z_k|X_{\mc{V}^c\cup\mc{S}^c}Q)\\
&-&I(\hat{Z}_{\mc{S}};{Z}_{\mc{S}}|X_{\mc{V}^c\cup\mc{T}_k}\hat{Z}_{\mc{S}^c}Z_kQ)\,\,\,\,\,\,\,\,\,\,\,\,(\mc{S}^c=\mc{T}_k-\mc{S}),
\end{IEEEeqnarray*}
for  $\mc{T},\mc{T}_k\subseteq \mc{V}\subseteq \mc{N}$ and $\mc{V}^c=\mc{N}-\mc{V}$. Moreover $\Upsilon(\mc{V})$ and $\Upsilon_k(\mc{V})$ are defined by\begin{align}
 \Upsilon(\mc{V})&=\{\mc{T}\subseteq\mc{V}: \text{ for all } \mc{S}\subseteq\mc{T}, Q_{\mc{T}}(\mc{S})\geq 0\}, \\
 \Upsilon_k(\mc{V})&=\{\mc{T}\subseteq\mc{V}: \text{ for all } \mc{S}\subseteq\mc{T}, Q^{(k)}_{\mc{T}}(\mc{S})\geq 0\},
\label{MCNNCcondition}
\end{align}
while $Q_{\mc{T}}(\mc{S})$ and  $Q^{(k)}_{\mc{T}}(\mc{S})$ are given by
\begin{IEEEeqnarray}{ll}
Q_{\mc{T}}(\mc{S})=I(X_{\mc{S}}&;\hat{Z}_{\mc{S}^c}Y_1|XX_{\mc{S}^c\cup\mc{V}^c}Q)\nonumber\\
&-I(Z_{\mc{S}};\hat{Z}_{\mc{S}}|XX_{\mc{T}\cup \mc{V}^c}\hat{Z}_{\mc{S}^c}Y_1Q), \\
Q^{(k)}_{\mc{T}}(\mc{S})=I(X_{\mc{S}}&;Z_k|X_{\mc{V}^c\cup\mc{S}^c}Q)\nonumber\\
&-I(\hat{Z}_{\mc{S}};Z_{\mc{S}}|XX_{\mc{V}^c\cup\mc{T}}\hat{Z}_{\mc{S}^c}Z_kQ).
\end{IEEEeqnarray}
\vspace{-2mm}
And $\mc{P}$ is the set of all admissible distributions:
\begin{align*}
\mathcal{P}= \displaystyle\Big \{P_{QXX_{\mc{N}}Z_{\mc{N}}\hat{Z}_{\mc{N}}Y_1}= P_{Q} &P_{XX_{\mc{V}^c}|Q} P_{Y_1Z_{\mc{N}}|XX_{\mc{N}}}\times\\
&\prod_{j\in \mc{V}} P_{X_j|Q}P_{\hat{Z}_j|X_jZ_jQ}\Big\}.  
\end{align*} 
\label{thm:3}
\end{theorem}
\vspace{-3mm}
Notice that $R_{\mc{T}}(\mc{S})$ (condition of correct decoding at the destination) is in general better than $R^{(k)}_{\mc{T}_k}(\mc{S})$ (condition of correct decoding at the relay $k$). This is because the destination uses backward decoding. Moreover using the same technique as \cite{Kramer2011}, it can be shown that the optimization in \eqref{CMNNC} can be done over $\mc{T}\subseteq\mc{V}$ instead of $\mc{T}\in\Upsilon(\mc{V})$. In other words the relays in $\mc{T}\subseteq\mc{V}^c$ can increase the rate only if they satisfy \eqref{MCNNCcondition}. It can be seen that for each $\mc{T}$, if there is $\mc{A}\subseteq\mc{T}$ such that $Q_{\mc{T}}(\mc{A})<0$, then we can remove the relays in $\mc{A}$ from $\mc{T}$ and the rate is improved by not using their compression, which is easier to manipulate in composite setting. By choosing $\mc{V}=\mc{N}$, the region of Theorem \ref{thm:3} is reduced to the same region, as in \cite{Wu2011,Kramer2011}, which is equivalent to NNC region \cite{Lim2011}. So Theorem \ref{thm:3} generalizes and includes the previous NNC scheme and it provides a potentially larger region. Indeed, for the degraded single relay channel, it is capacity achieving while NNC is strictly suboptimal. In fact, relay nodes are divided into two groups. The first group is $\mc{V}^c$ which are using DF coding and the second group $\mc{V}$ which are using CF scheme.

Now consider the composite setting again. Before starting the communication, the source knows that the relays with index in $\mc{V}$ use CF while the others use DF scheme. For each $\uptheta$ and $\mc{V}$, the rate $r$ is achievable if it belongs to the region in the previous theorem and otherwise an error is declared. All that CF relays can do in this case is to choose their distribution based on $\uptheta_r$ such that it minimizes the error probability. Thus the expected error probability for the composite multiple relay channel with partial CSI $\theta_r$ at the relays is bounded by
\begin{IEEEeqnarray*}{ll}
\bar{\epsilon}(r) &\leq \inf_{\mc{V}\subseteq\mc{N}}\min_{p(x,x_{\mc{V}^c},q)}\nonumber\\
&\prob[E]_{\uptheta_r} 
\Big\{
\min_{\prod_{j\in \mc{V}}p({x_j|q})p({\hat{z}_j|x_jz_jq})}  
\prob_{\uptheta|\uptheta_r}\big[r > I_{\textrm{CMNNC}}(\mc{V}) \big| \uptheta_r\big] 
\Big\}
\label{CMNNCoutage}   
\end{IEEEeqnarray*}
where for all $\uptheta=(\uptheta_r,\uptheta_d)$, $I_{\textrm{MNNC}}(\mc{V})=$
\begin{align}
&\min\Big\{\max_{\mc{T}\subseteq\mc{V}}\min_{\mc{S}\subseteq\mc{T}} R_{\mc{T}}(\mc{S},\uptheta), \min_{k\in\mc{V}^c}\max_{\mc{T}_k\in\Upsilon_k(\mc{V})}\min_{\mc{S}\subseteq\mc{T}_k} R^{(k)}_{\mc{T}_k},(\mc{S},\uptheta)\Big\},
\label{IV}
\end{align}\vspace{-2mm}
% \begin{IEEEeqnarray}{ll}
% I_{\textrm{MNNC}}(\mc{V})=\min&\Big\{\max_{\mc{T}\subseteq\mc{V}}\min_{\mc{S}\subseteq\mc{T}} R_{\mc{T}}(\mc{S},\uptheta), \nonumber\\
% & \min_{k\in\mc{V}^c}\max_{\mc{T}_k\in\Upsilon_k(\mc{V})}\min_{\mc{S}\subseteq\mc{T}_k} R^{(k)}_{\mc{T}_k}(\mc{S},\uptheta)\Big\}. 
% \label{IV}
% \end{IEEEeqnarray}
\begin{align*}
R_{\mc{T}}(\mc{S},\uptheta)&=I(XX_{\mc{V}^c}X_{\mc{S}};\hat{Z}_{\mc{S}^c}Y_{1\uptheta}|X_{\mc{S}^c}Q)\nonumber\\
&-I(Z_{\mc{S}\uptheta_r};\hat{Z}_{\mc{S}}|XX_{\mc{T}\cup \mc{V}^c}\hat{Z}_{\mc{S}^c}Y_{1\uptheta}Q) (\mc{S}^c=\mc{T}-\mc{S}), \nonumber\\
R^{(k)}_{\mc{T}_k}(\mc{S},\uptheta)&=I(X;\hat{Z}_{\mc{T}_k}Z_{k\uptheta_r}|X_{\mc{V}^c}X_{\mc{T}_k}Q) \nonumber\\
+I(X_{\mc{S}};&Z_{k\uptheta_r}|X_{\mc{V}^c\cup\mc{S}^c}Q)-I(\hat{Z}_{\mc{S}};{Z}_{\mc{S}\uptheta_r}|X_{\mc{V}^c\cup\mc{T}_k}\hat{Z}_{\mc{S}^c}Z_{k\uptheta_r}Q) 
\end{align*}
for  $\mc{T},\mc{T}_k\subseteq \mc{V}\subseteq \mc{N}$ and $\mc{V}^c=\mc{N}-\mc{V}$. Similarly $\Upsilon_k(\mc{V})$ is 
\vspace{-1mm}
\begin{align}
\Upsilon_k(\mc{V})&=\{\mc{T}\subseteq\mc{V}: \text{ for all } \mc{S}\subseteq\mc{T}, Q^{(k)}_{\mc{T}}(\mc{S},\uptheta_r)\geq 0\}
\label{CMNNCrates}
\end{align}
\vspace{-1mm}
where $Q^{(k)}_{\mc{T}}(\mc{S},\uptheta_r)$ is defined as follows:
\begin{align*}
Q^{(k)}_{\mc{T}}(\mc{S},\uptheta_r)=&I(X_{\mc{S}};Z_{k\uptheta_r}|X_{\mc{V}^c\cup\mc{S}^c}Q)\\
&-I(\hat{Z}_{\mc{S}};Z_{\mc{S}\uptheta_r}|XX_{\mc{V}^c\cup\mc{T}}\hat{Z}_{\mc{S}^c}Z_{k\uptheta_r}Q).
\end{align*}
% \vspace{-1mm}
In the preceding scheme all relays fix coding regardless of the available CSI. However, it is possible that the relays select and change their coding based on its CSI. To this purpose each relay generates many codebooks and sends one of them which fits the best to the channel with the parameter $\uptheta_r$. More precisely, each relay $k$ has a decision region $\mc{D}^{(k)}_{\textrm{DF}}$ such that for all $\theta_r\in\mc{D}^{(k)}_{\textrm{DF}}$, the relay $k$ uses DF scheme and otherwise it uses CF scheme.  For each $\mc{V}\subseteq\mc{N}$, define $\mc{D}_{\mc{V}}$ as follows:
$$
\displaystyle\mc{D}_{\mc{V}}=\left(\bigcap_{k\in\mc{V}^c}{\mc{D}^{(k)}_{\textrm{DF}}}\right) \bigcap \left(\bigcap_{k\in\mc{V}}{\mc{D}^{(k)}_{\textrm{DF}}}^c\right).
$$
If $\theta_r\in\mc{D}_{\mc{V}}$, then $\theta_r\notin\mc{D}^{(k)}_{\textrm{DF}}$ for all $k\in\mc{V}$, and $\theta_r\in\mc{D}^{(k)}_{\textrm{DF}}$ for all $k\notin\mc{V}$. So the  $k$-th relay, for each $k\in\mc{V}$ uses CF and the relay $k'$ for $k'\in\mc{V}^c$ uses DF. The ensemble of decision regions of relays will thus provide the regions $\mc{D}_{\mc{V}}$ which are mutually disjoint and all together form a partitioning over the set $\Theta_r$. Now if $\theta_r\in\mc{D}_{\mc{V}}$, we have a multiple relay network where the relays in $\mc{V}$ are using CF. The achievable rate corresponding to this case is known from Theorem \ref{thm:3}.

As shown in Fig. \ref{fig:CMRN}, each relay has two set of codewords: $X^{(1)}_{(k)}$ and $X^{(2)}_{(k)}$. The first code $X^{(1)}_{(k)}$ is transmitted  when $\theta_r\in\mc{D}^{(k)}_{\textrm{DF}}$. This code is based on DF strategy so the $k$-th relay decodes the source message and transmits it to the destination. However the source, not knowing whether the  $k$-th relay is sending $X^{(1)}_{(k)}$ or not, uses superposition coding and superimpose its code over $X^{(1)}_{(k)}$. If the $k$-th relay sends $X^{(1)}_{(k)}$ then this will become DF relaying. The source, oblivious to the relays decision, generates its own code by using superposition coding and $X$ is superimposed over $X^{(1)}_{\mc{N}}$, i.e., all possible DF relay inputs. This does not affect $R_{\mc{T}}(\mc{S},\uptheta)$ by applying the proper Markov chain, but changes $R^{(k)}_{\mc{T}_k}(\mc{S},\uptheta)$, as we will see.
% The optimum correlation between the source and the relay cannot be adaptively found based on $\theta_r$.  

On the other hand, if $\theta_r\notin\mc{D}^{(k)}_{\textrm{DF}}$ then CF scheme is used. Note that unlike DF, the code which is used for CF $X^{(2)}_{(k)}$, is independent of the source code and so its PD can be chosen adaptively based on $\theta_r$. The optimum choice for $\mc{D}_{\mc{V}}$ will potentially give a better outage probability than the case that each relay is using a fixed coding for all $\theta_r$. This provides a non-formal proof of the next theorem.\vspace{1mm}

\begin{theorem} [SCS with partial CSI-cooperative relays]
\label{thm:5}
The average error probability of the composite multiple relay channel with partial CSI $\theta_r$ at the relays can be upper bounded by 
$\bar{\epsilon}(r) \leq \min_{p(x,x^{(1)}_{\mc{N}},q)}\inf_{\left\{\mc{D}_{\mc{V}},\mc{V}\subseteq\mc{N}\right\}\in\Pi\left(\Theta_r,N\right)}$
\begin{align}
\displaystyle\sum_{\mc{V}\subseteq\mc{N}}
\prob[E]_{\uptheta_r} 
\Big\{
&\min_{\prod_{j\in \mc{V}}p({x^{(2)}_j|q})p({\hat{z}_j|x^{(2)}_jz_jq})}  \nonumber\\
&\prob_{\uptheta|\uptheta_r}\big[r > I_{\textrm{CMNNC}}(\mc{V}),\uptheta_r \in \mc{D}_{\mc{V}} \big| \uptheta_r\big] \Big\}, 
\label{SCSoutage-CMR-1}   
\end{align}
$\Pi\left(\Theta_r,N\right)$ is the set of all partitioning over $\Theta_r$ into at most $2^N$ disjoint sets. The relay inputs  $X_{k}$ is chosen from $(X^{(1)}_{k},X^{(2)}_{k})$ such that $X_{k}$ is equal to $X^{(1)}_{k}$ if $\displaystyle\theta_r\in\mc{D}^{k}_{\textrm{DF}}$ and equal to $X^{(2)}_{k}$ if $\displaystyle\theta_r\notin\mc{D}^{k}_{\textrm{DF}}$.
% ($\displaystyle\mc{D}^{k}_{\textrm{DF}}=\bigcup_{\mc{V}\subset\mc{N},k\notin\mc{V}}\mc{D}_{\mc{V}}$). 
Indeed, for $\uptheta_r\in\mc{D}_{\mc{V}}$ the next Markov chain holds: $$(X^{(1)}_{\mc{V}},X^{(2)}_{\mc{V}^c}) \minuso (X,X^{(1)}_{\mc{V}^c},X^{(2)}_{\mc{V}}) \minuso (Y_{1\uptheta},Z_{\mc{N}\uptheta_r}),$$ 
where $I_{\textrm{MNNC}}(\mc{V})$ and $\Upsilon_k(\mc{V})$ are defined by expressions \eqref{IV} and \eqref{CMNNCrates} with the difference that in $R^{(k)}_{\mc{T}_k}(\mc{S},\uptheta)$ and $Q^{(k)}_{\mc{T}}(\mc{S},\uptheta_r)$, $X_{\mc{V}^c}$ is replaced with $X^{(1)}_{\mc{N}}$.  
\end{theorem}

\section{Sketch of the Proof of Theorem \ref{thm:3}}
Consider first the two relay network.  Relay $1$ uses DF scheme to help the source so it has to decode the source messages successively and not backwardly, and Relay $2$ uses CF scheme. However, relay $1$ wants to exploit the help of relay $2$ to decode the source message. So it does not start decoding until it retrieves the compression index. To this end, relay $1$ uses offset decoding which means that it waits two blocks instead of one to decode the source message and the compression index. In block $b=2$, the relay $1$ decodes the compression index $l_1$ and the message $w_1$. Equally, the source code at block $b+2$ is correlated with relay $1$ code from the block $b$ and not block $b+1$. This comes at the expense of one block of delay. The source has to wait until $b=B+L$ to start backward-decoding. The compression index $l_{B+2}$ is repeated until the block $B+L$. Fix $P$, $\mc{V},\mc{T}$ and $\mc{T}_k$'s  such that they maximize the right hand 
side of \eqref{CMNNC}. Assume a 
set $\mc{M}_n$ of size $2^{nR}$ of message indices $W$ to be transmitted, again in $B+L$ blocks, each of them of length $n$. At the last $L-2$ blocks, the last compression index is first decoded and then all compression indices and transmitted messages are jointly decoded. Relays in $\mc{V}^c$ start to decode after block $2$. \vspace{2mm}\\
\textit{Code generation:}
\begin{enumerate} [(i)]
\item Randomly and independently generate $2^{nR}$ sequences $\underline{x}_{\mc{V}^c}$ drawn i.i.d. from $P_{X_{\mc{V}^c}}^{n}(\underline{x}_{\mc{V}^c})=\prod\limits_{j=1}^n P_{X_{\mc{V}^c}}(x_{{\mc{V}^c}j}).$  Index them as $\underline{x}_{\mc{V}^c}(r)$ with index $r\in \left[1,2^{nR}\right]$. 
\item For each $\underline{x}_{\mc{V}^c}(r)$, randomly and conditionally independently generate $2^{nR}$ sequences $\underline{x}$ drawn i.i.d. from  $P_{X|X_{\mc{V}^c}}^n(\underline{x}\vert \underline{x}_{\mc{V}^c}(r))=\prod\limits_{j=1}^n P_{X|X_{\mc{V}^c}}(x_{j}|x_{{\mc{V}^c}j}).$ Index them as $\underline{x}(r,w)$, where $w\in \left[1,2^{nR}\right]$.
\item For each $k\in\mc{V}$, randomly and independently generate  $2^{n\hat{R}_k}$ sequences $\underline{x}_k$ drawn i.i.d. from  $P_{X_k}^n(\underline{x}_k)=\prod\limits_{j=1}^n P_{X_k}(x_{kj}).$ 
Index them as $\underline{x}_k(r_k)$, where $r_k\in [1,2^{n\hat{R}_k}]$ for $\hat{R}_k=I(Z_k;\hat{Z}_k|X_k)+\epsilon$.
\item For each $k\in\mc{V}$ and each $\underline{x}_k(r_k)$, randomly and conditionally independently generate $2^{n\hat{R}_k}$ sequences  $\underline{\hat{z}}_k$ each with probability $P_{\hat{Z}_k|X_k}^n (\underline{\hat{z}}_k\vert \underline{x}_k(r_k))= \prod\limits_{j=1}^n P_{\hat{Z}_k\vert X_k}(\hat{z}_{kj}\vert x_{kj}(r_k)).$ Index them as  $\underline{\hat{z}}_k(r_k,\hat{s}_k)$, where $\hat{s}_k\in [1,2^{n\hat{R}_k}]$.\\
\end{enumerate}
%%%%%%%%%%%%%%%%%%%%%%%%%%%%%%%%%%%%%%%%%%%%%%%Encoding Part
\textit{Encoding part:} 
\begin{enumerate}[(i)]
\item
In every block $i=[1: B]$, the source sends $w_{i}$ using $\underline{x}\big(w_{(i-2)},w_i\big)$ ($w_{0}=w_{-1}=1$). Moreover, for blocks $i=[B+1:B+L]$, the source sends the dummy message $w_{i}=1$ known to all users.
\item
For every block $i=[1:B+L]$, and each $k\in\mc{V}^c$, the relay $k$ knows $w_{(i-2)}$ by assumption and $w_0=w_{-1}=1$, so it sends $\underline{x}_k\big(w_{(i-2)}\big)$. 
\item 
For each $i=[1: B+2]$, each $k\in\mc{V}$, the relay $k$ after receiving $\underline{z}_k(i)$, searches for at least one index $l_{ki}$ with $l_{k0}=1$ such that 
$$
\big(\underline{x}_{k}(l_{k(i-1)}),\underline{z}_{k}(i),\underline{\hat{z}}_{k}(l_{k(i-1)},l_{ki})\big)\in \mc{A}^n_\epsilon[X_kZ_k\hat{Z}_k].
$$ 
The probability of finding such $l_{ki}$ goes to one as $n$ goes to infinity due to the choice of $\hat{R}_k$. 
\item
For $i=[1:B+2]$ and $k\in\mc{V}$, relay $k$ knows from the previous block ${l}_{k(i-1)}$ and it sends $\underline{x}_k(l_{k(i-1)})$. Moreover, relay $k$ repeats $l_{k(B+2)}$ for $i=[B+3: B+L]$, which means for $L-2$ blocks.
\end{enumerate}\vspace{1mm}
%%%%%%%%%%%%%%%%%%%%%%%%%%%%%%%%%%%%%%%%%%%%%%%Decoding Part
\textit{Decoding part:} 
\begin{enumerate}[(i)]
\item After the transmission of the block  $i=[1:B+1]$ and for each $k\in\mc{V}^c$, with the assumption that all messages and compression indices up to block $i-1$ have been correctly decoded, the $k$-th relay  searches for the unique index $(\hat{w}_{b},\hat{l}_{\mc{T}_kb})$ by looking at two consecutive blocks $b$ and $b+1$ such that:
\begin{align*}
&\Big(\underline{x}({w}_{(b-2)},\hat{w}_{b}),\underline{x}_{\mc{V}^c}({w}_{(b-2)}),\underline{z}_{k}(b),
\Big(\underline{x}_{k}({l}_{k(b-1)}),\\
&\underline{\hat{z}}_{k}({l}_{k(b-1)},\hat{l}_{kb})\Big)_{k\in\mc{T}_k}\Big)\in \mc{A}^n_\epsilon[XX_{\mc{T}_k\cup\mc{V}^c}\hat{Z}_{\mc{T}_k}Z_k] \text{ and } \\
&\big(\underline{x}_{\mc{V}^c}({w}_{(b-1)}),\underline{z}_{k}(b+1),(\underline{x}_{k}(\hat{l}_{kb}))_{k\in\mc{T}_k}\big)\in \mc{A}^n_\epsilon[X_{\mc{T}_k\cup\mc{V}^c}Z_k].
\end{align*}
Probability of error goes to zero as $n\to\infty$ if

\begin{align}
R \leq& I(X;\hat{Z}_{\mc{T}}Z_k|X_{\mc{V}^c}X_{\mc{T}_k})+I(X_{\mc{S}};Z_k|X_{\mc{V}^c}X_{\mc{S}^c}) \nonumber\\
&-I(\hat{Z}_{\mc{S}};{Z}_{\mc{S}}|X_{\mc{V}^c\cup\mc{T}_k}\hat{Z}_{\mc{S}^c}Z_k). \label{DFdecoding-CMNNC}\\
0 \leq& I(Z_k;X_{\mc{S}}|X_{\mc{V}^c\cup\mc{S}^c})-I(\hat{Z}_{\mc{S}};Z_{\mc{S}}|XX_{\mc{V}^c\cup\mc{T}_k}\hat{Z}_{\mc{S}^c}Z_k). \label{Compression-relay-CMNNC}
\end{align}

Given the fact that $\mc{T}_k\in\Upsilon_k(\mc{V})$, the last inequality holds for each $\mc{S}\subseteq\mc{T}_k$. 
\item
The destination jointly searches for the unique indices $(\hat{l}_{k(B+2)})_{k\in\mc{T}}$ such that for all $b\in[B+3:B+L]$,
$\Big((\underline{x}_{k}(\hat{l}_{k(B+1)}))_{k\in\mc{T}},\underline{x}(1,1),\underline{x}_{\mc{V}^c}(1),\underline{y}_{1}(b)\Big)$ belongs to $\mc{A}^n_\epsilon[XX_{\mc{T}}X_{\mc{V}^c}Y_1].$
The probability of error goes to zero as $n\to\infty$ provided for all subsets $\mc{S}\subseteq\mc{T}$:
\begin{align*} 
&\sum_{k\in\mc{S}} I(\hat{Z}_k;Z_k|X_k)\leq (L-2)I(X_{\mc{S}};XX_{\mc{S}^c\cup\mc{V}^c}Y_1).
\label{Compression-1-CMNNC}
\end{align*}
\item 
With the assumption that $(w_{b+2},l_{\mc{T}(b+2)})$ have been correctly decoded, the destination finds backwardly the unique pair of indices $(\hat{w}_b,\hat{l}_{\mc{T}(b+1)})$ such that:
\begin{align*}
&\Big(\underline{x}(\hat{w}_{b},w_{(b+2)}),\underline{x}_{\mc{V}^c}(\hat{w}_{b}),\underline{y}_{1}(b+2),
\big(\underline{x}_{k}(\hat{l}_{k(b+1)}),\\
&\underline{\hat{z}}_{k}(\hat{l}_{k(b+1)},l_{k(b+2)})\big)_{k\in\mc{T}}\Big)\in \mc{A}^n_\epsilon[XX_{\mc{T}\cup\mc{V}^c}\hat{Z}_{\mc{T}}Y_1].
\end{align*}  
The probability of error goes to zero as $n\to\infty$ if:
\begin{align*}
0  < I(\hat{Z}_{\mc{S}^c}Y_1;X_{\mc{S}}&|XX_{\mc{V}^c\cup\mc{S}^c})\\ &- I(\hat{Z}_{\mc{S}};Z_{\mc{S}}|XX_{\mc{V}^c\cup\mc{T}}\hat{Z}_{\mc{S}^c}Y_1), \\%\label{Compression-2-CMNNC} 
R  \leq I(XX_{\mc{V}^c}X_{\mc{S}}&;Y_1\hat{Z}_{\mc{S}^c}|X_{\mc{S}^c})\\&-I(\hat{Z}_{\mc{S}};{Z}_{\mc{S}}|XX_{\mc{V}^c\cup\mc{T}}\hat{Z}_{\mc{S}^c}Y_1).
%\label{CFdecoding-2-CMNNC}
\end{align*}
Using the previous inequalities, and by choosing finite $L$ but large enough, by letting $(B,n)\to\infty$ and adding the time-sharing random variable $Q$ the proof is finished. 
\end{enumerate}
%%%%%%%%%%%%%%%%%%%%%%%%%%%%%%%%%%%%%%%%%%%%%%%%%%%%%%%%%%%%%%%%%%%%%%%%%%%%%%%%%%%%%%%%%%%%%%%%%%%%%%%%%%%%%%%%%%%%%%%%%%
%%%%%%%%%%%%%%%%%%%%%%%%%%%%%%%%%%%%%%%%%%%%%%%%%%%%%%%%%%%%%%%%%%%%%%%%%%%%%%%%%%%%%%%%%%%%%%%%%%%%%%%%%%%%%%%%%%%%%%%%%%
%% Guassian Slowly Fading Networks
%%%%%%%%%%%%%%%%%%%%%%%%%%%%%%%%%%%%%%%%%%%%%%%%%%%%%%%%%%%%%%%%%%%%%%%%%%%%%%%%%%%%%%%%%%%%%%%%%%%%%%%%%%%%%%%%%%%%%%%%%%
%%%%%%%%%%%%%%%%%%%%%%%%%%%%%%%%%%%%%%%%%%%%%%%%%%%%%%%%%%%%%%%%%%%%%%%%%%%%%%%%%%%%%%%%%%%%%%%%%%%%%%%%%%%%%%%%%%%%%%%%%%
\section{Gaussian Slow-Fading Networks}
\begin{figure} [t]
\centering  
\includegraphics [width=.45 \textwidth] {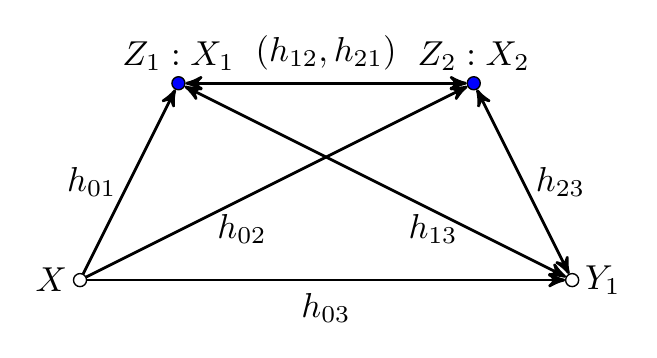}    
\caption{Fading Gaussian two-relay channel.}  \vspace{-4mm}
\label{fig:2relay}
\end{figure}
Consider the Gaussian fading two-relay network, depicted in Fig. \ref{fig:2relay}, which is defined by the following relations:
\begin{align*}
Z_1& = \frac{h_{01}}{d^\alpha}X+h_{21}X_2+ \mc{N}_1,\\
Z_2& = h_{02}X+h_{12}X_1+ \mc{N}_2,\\
Y_1& = h_{03}X+h_{13}X_1+h_{23}X_2+\mc{N}_3.
\end{align*}
Define $\mc{N}_i$'s to be additive noises, i.i.d. circularly symmetric complex Gaussian RVs with zero-mean; let $h_{ij}$'s be independent zero-mean circularly symmetric complex Gaussian RVs. 
% Set the fading matrix $\mathbf{H}$, and 
$d$ is the random path-loss.  The average power of the source and relay inputs $X$, $X_1$ and $X_2$ must not exceed powers $P$, $P_1$ and $P_2$, respectively. Compression is obtained by adding additive noises $\hat{Z}_1 = Z_1+\hat{\mc{N}}_1$, $\hat{Z}_2 = Z_2+ \hat{\mc{N}}_2.$ It is assumed that the source is not aware of the fading coefficients, the relays know all fading coefficients except $h_{i3}$'s and the destination is fully aware of everything. The possibilities to choose the proper cooperative strategy are as follows: (i) both relays use DF scheme to transmit the information (full DF case), (ii) both relays use CF scheme to transmit the information (full CF case), where the destination can consider one or both relays as noise to prevent the performance 
degradation, and (iii) one relay uses DF scheme and the other uses CF scheme (Mixed Coding case). Finally, the relays can select their coding strategy based on the channel parameters (SCS case). 

\begin{figure} [t]
\centering  
\includegraphics [width=.45 \textwidth] {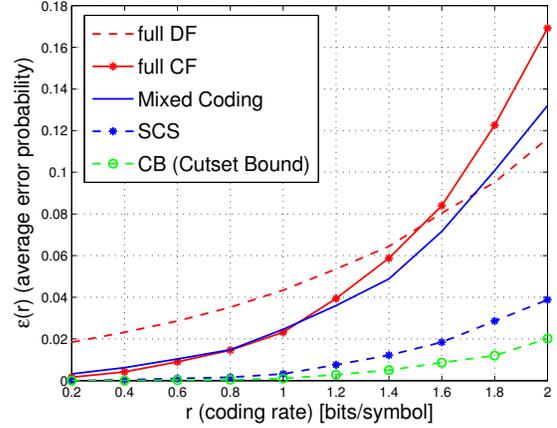}    
\caption{Asymptotic error probability $\bar{\epsilon}(r)$ vs. coding rate $r$.}
\label{fig:2}
\end{figure}

Fig. \ref{fig:2} presents  numerical analysis of these strategies. We assume all fading coefficients are of unit variance and so are the noises. $d$ is chosen with uniform distribution between 0 and 0.1, which means the first relay is always around the source. Given this assumption, we suppose that the first relay uses DF in case of mixed coding while the other uses CF scheme.  The source and relay powers are respectively 1 and 10. It can be seen that none of the non-selective strategies like full DF, full CF and Mixed Coding is not in general the best regardless of fading coefficients. However, if one lets the relay select their strategy given its channel measurement, this SCS will lead to significant improvement compared to the other strategies and becomes close to the cutset bound. 
% In this setting the relays exploit efficiently their CSI to adapt the coding to the channel. t
% 
% \section{Conclusion}
% In this paper, the composite multiple relay network was discussed where the channel is randomly drawn according to $\prob_\uptheta$ and it is assumed to be unknown at the source, fully known at the destination and only partly known at the relays. A novel coding was introduced for this channel enabling the relays to select the best coding strategy based on their CSI. It was shown that Noisy Network Coding theorem can be used in case of mixed strategy with DF and CF relays where DF  relays benefit from the help of CF relays using offset coding. It was shown via numerical results that Selective Coding Strategy (SCS) improves significantly other schemes. The future works consists in generalizing this scheme to multicast setting.
% % use section* for acknowledgement
% \section*{Acknowledgment}
\bibliographystyle{IEEEtran}
\bibliography{biblio}

\end{document}